\documentclass[preprint,11pt,reqno]{amsart}
\pdfoutput=1
\usepackage{graphicx}
\usepackage{amsmath,amstext,amsbsy,amssymb,color,caption}
\usepackage[cdot,amssymb]{SIunits}
\usepackage[latin1]{inputenc}
\usepackage{bm}
\DeclareMathOperator{\sign}{Sign} 

\title[]{Preprint\\\vspace{2\baselineskip} \emph{THE INFLUENCE OF DEFLECTIONS ON THE STATIC AND DYNAMIC BEHAVIOUR OF MASONRY
COLUMNS}}

\author{MARIA GIRARDI}
\address{Istituto di Scienza e Tecnologie dell'Informazione "A. Faedo", CNR
Via G. Moruzzi 1\\
Pisa, 56124\\
Italy} \email{Corresponding Author: Maria.Girardi@isti.cnr.it}

\author{CRISTINA PADOVANI}

\author{DANIELE PELLEGRINI}


\def\thesubsection{\thesection\Alph{subsection}}
\numberwithin{equation}{section}
\def\theequation{\thesection\hbox{-}\arabic{equation}}

\begin{document} 

\begin{abstract}

This paper studies the influence of bending deflections on the
structural behaviour of masonry columns. Some explicit solutions are
presented, and the combined effects of the constitutive and
geometric nonlinearities are investigated through an iterative
numerical procedure. The results show that considering second-order
effects affects both the collapse load and the dynamical properties
of masonry beams significantly.
\end{abstract}

\maketitle 

\section{Introduction}\label{sec:sec1}

Masonry buildings are unable to withstand loads with large
eccentricities. Ancient masonry constructions are mainly designed to
constrain the compressive force inside the elements' section, while
large tensile stresses are concentrated in the wooden and metallic
parts. On the other hand, bending is always present in masonry
elements. When the axial force is applied outside the central
nucleus of inertia of a masonry beam, there is a reduction of the
section's stiffness, and the behaviour of the beam becomes
nonlinear. Many constitutive equations have been proposed during the
last century to describe the peculiar behaviour of unreinforced
masonry materials, which are essentially unable to withstand tensile
stresses (see \cite{COMO}, \cite{SPRING} for a review). More
recently, the problem of bending in masonry has been addressed by
several authors, mainly in the framework of earthquake engineering,
being masonry constructions prone to seismic actions.


When deformation is taken into account in the equilibrium equations,
geometric and constitutive nonlinearities are coupled, and the
effects of bending are consequently amplified. First investigations
on the stability of masonry pillars date back to the Seventeenths,
with the studies \cite{yokel1971}, \cite{frisch1975}, and
\cite{frisch1980}, and successively recalled in \cite{ADF2002},
\cite{ADF2003}, \cite{DeFalco2007}. In all these studies, some
explicit solutions are proposed to determine the collapse load of
masonry pillars subjected to eccentric loads. In \cite{la1993},
\cite{popehn2008} some iterative procedures are shown to evaluate
the effects of deformation on the equilibrium of simple masonry
elements;  \cite{popehn2008} also presents the results  of an
experimental campaign on masonry panels subjected to transverse and
axial loads. In \cite{Pin2007}, a finite--element analysis is
proposed to evaluate the equilibrium of masonry beams in the
presence of geometric nonlinearities.  The effects of large
deformations, together with those of the construction phases, are
taken into account in the finite--element analysis conducted in
\cite{pela} to assess the static conditions of the Mallorca
cathedral.

The correlation between changes in the natural frequencies and the
presence of structural damage \cite{Salawu}, \cite{Review2021},
\cite{Modal}, confirmed by many dynamic monitoring campaigns
\cite{Ref1}, makes it interesting to investigate how cracks affect
the dynamic behaviour of structures. As far as slender structures
are concerned, the influence of fractures on the modal properties of
beams has been addressed and modelled in several papers, and a
comprehensive list of references is reported in \cite{Gurel}.

This paper investigates the influence of geometric nonlinearity on
the static and dynamic behaviour of Euler-Bernoulli beams made of a
masonry-like material \cite{GirardiandLucchesi}. The invertibility
of the moment-curvature function allows determining the explicit
expression of the transverse displacement in masonry cantilever
beams subjected to prescribed forces applied at the free end. Two
cases are addressed: in the former (case (a)) the beam is subjected
to an eccentric normal load N, in the latter (case (b)) the axial
force N is applied along with a horizontal load H. The knowledge of
the normal force and bending moment along the beam's axis makes it
possible to calculate the deflection while considering both material
and geometric nonlinearities. When second-order effects are taken
into account, the nonlinear differential equation linking deflection
and curvature is integrated via an iterative scheme, thus providing
response curves analogous to those available in the literature
\cite{ADF2002}, \cite{yokel1971}. The results of the numerical
approach proposed in Section 2 are compared with those of
finite--element analyses conducted with the Marc code using a
concrete cracking model for masonry \cite{marc}. Section 3 is
devoted to assessing the influence of geometric nonlinearity on the
natural frequencies of masonry beams. The fundamental frequency of
simply supported beams subjected to two different load conditions is
calculated explicitly by using the results of Section 2. The
dependence of the frequency on the loads is validated via the Marc
code and plots of the frequency vs the eccentricity of load N and
the horizontal load H are provided.

\section{Some explicit solutions}\label{sec:sec2}

Let us consider a rectilinear beam with a rectangular cross--section
of height $h$ and width $b$, subjected to an axial force $N<0$. Let
us denote by $E$ the Young's modulus  and by $J = bh^3/12$ the
moment of inertia of the beam's cross section, and let $x$ be the
abscissa along the beam's axis, $y(x)$ the beam's transverse
deflection at $x$. The beam is  modelled according to the
Euler-Bernoulli beam theory.

The curvature of the beam's deflection is denoted by $\chi$ and,
under the hypothesis that the rotations of the beam's axis are
small, is given by

\begin{equation}
\chi(x)=-\frac{d^2 y}{d x^2},\label{eq:eq3}
\end{equation}
where $\frac{d^2 y}{d x^2}$ denotes the second derivative with
respect to $x$.

The bending moment $M(\chi)$ is a continuously differentiable
function of $\chi$, whose second derivative is assumed to be
piecewise continuous.

For a beam constituted by a masonry--like material with infinite
compressive strength and zero tensile strength
\cite{GirardiandLucchesi}, function $M(\chi)$ is

\begin{equation}
\label{eq:eq1} M(\chi)=
\begin{cases}
\quad EJ\,\chi \quad &\text{for $\left| \chi \right|
\le \alpha$},\\
\quad  EJ\, \alpha   \sign(\chi )  (3 - 2\sqrt {\frac{\alpha
}{{\left| \chi  \right|}}} )\quad &\text{for} \left| \chi \right|
> \alpha,
\end{cases}
\end{equation}
where

\begin{equation}\label{eq:eq2}
\alpha=-\frac{2N}{Ebh^2}
\end{equation}
is the curvature corresponding to the elastic limit.

\begin{figure}
\centering
\includegraphics[width=13cm]{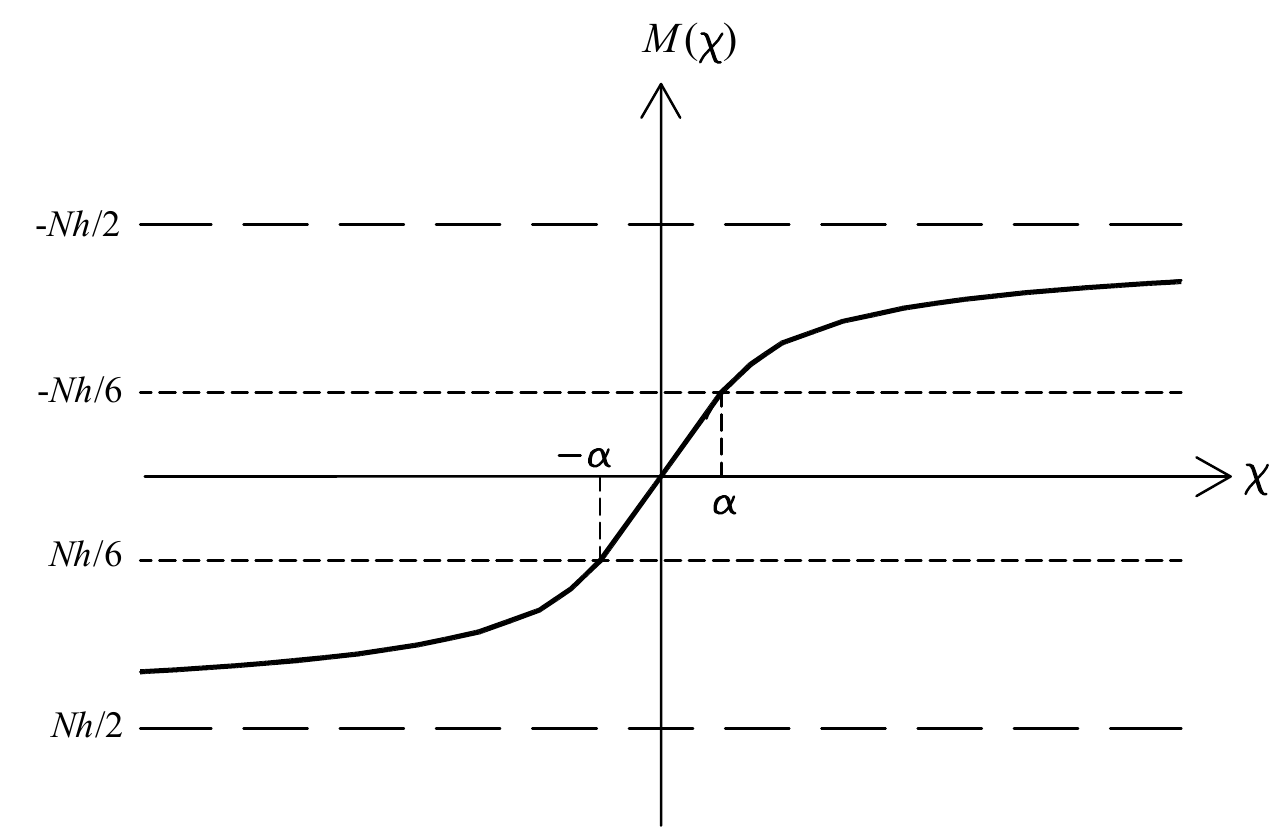}
\caption{Bending moment $M$ versus the curvature $\chi$ for a
rectangular cross--section beam  with infinite compressive strength
and zero tensile strength.}\label{fig:fig8}
\end{figure}

Function $M(\chi)$ depicted in Figure \ref{fig:fig8} is invertible
and its inverse is

\begin{equation}
\label{eq:eq5a} \left|\chi(M)\right|=
\begin{cases}
\quad \frac{\left|M\right|}{EJ} \quad &\text{for $ \left|M\right|
\le - N\,h/6$},\\
\vspace{1\baselineskip} \\
4 \,\alpha^3/\left(\frac{\left|
M\right|}{EJ}-3\,\alpha\right)^2\quad &\text{for\, $ \left|M\right|
>-N\,h/6$}.
\end{cases}
\end{equation}

\begin{figure}
\centering
\includegraphics[width=12cm]{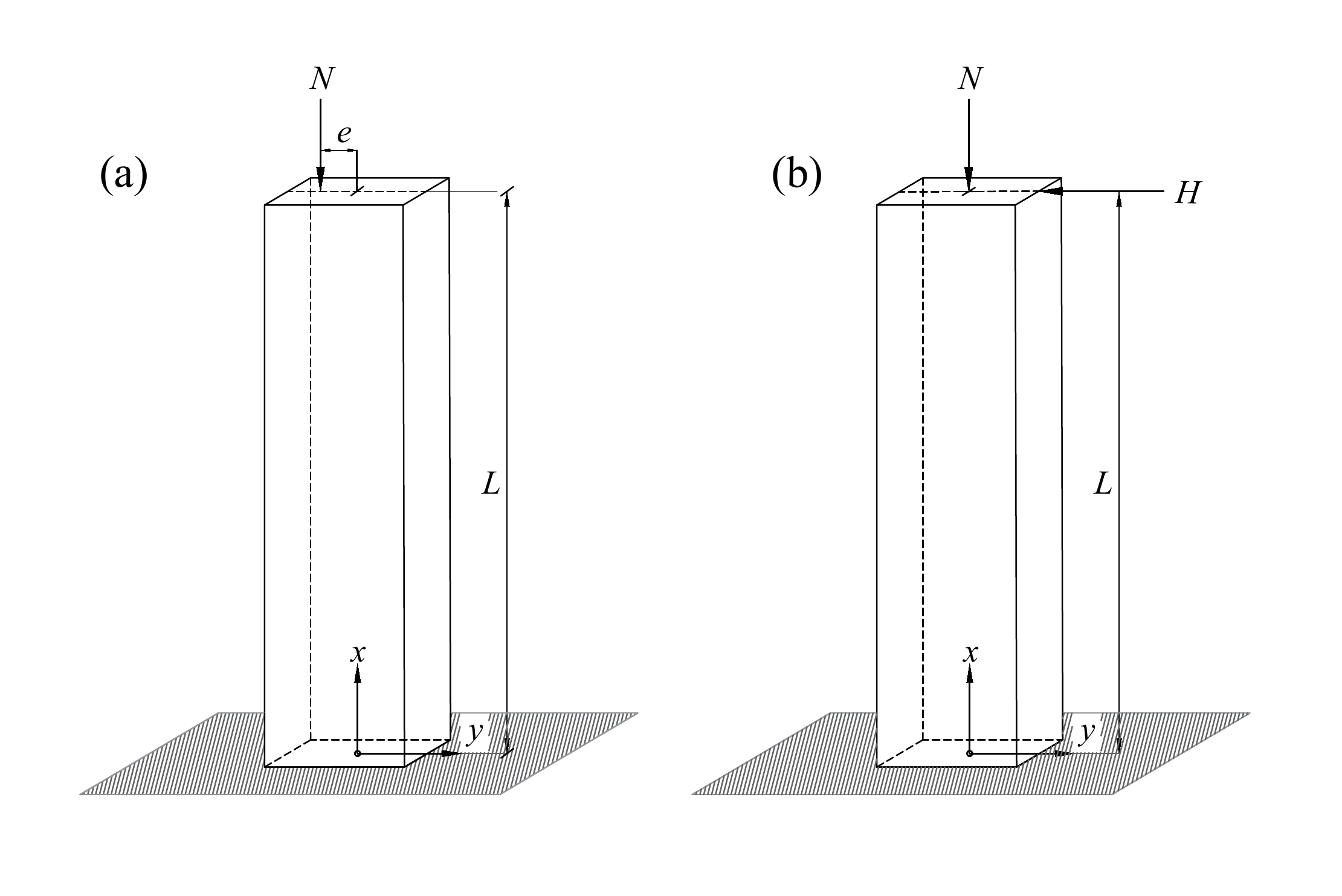}
\caption{A cantilever beam. Case (a): eccentric axial load. Case
(b): horizontal and axial loads.}\label{mensola}
\end{figure}

Under the hypotheses above, if the bending moment $M$ and the normal
force $N$ acting on the beam's sections are known along the axis,
the deflection of the beam can be easily calculated by integrating
equation \eqref{eq:eq3} with the help of \eqref{eq:eq5a}. Thus, a
number of explicit solutions can be obtained, some of which are
reported in the following.

Let us consider the cantilever beam with length $L$ represented in
Figure \ref{mensola}. In case (a) the beam is subjected to a normal
force $N$ acting on the beam's upper end with eccentricity $e\ge0$.
Neglecting the second--order effects, that is, ignoring the
contribution of displacements and rotations on the beam's
equilibrium, the eccentricity of the axial load remains constant
along the beam and, by \eqref{eq:eq5a}, curvature $\chi$ is constant
as well. In case (b), the beam is subjected to the vertical load $N$
and a horizontal load $H$, both acting on the upper end. Both cases,
(a) and (b), are relevant in many applications concerning the static
behaviour of masonry pillars. In particular, case (b)  is adopted to
model the seismic actions through the so--called \emph{push--over}
analysis. Finally, it is worth noting that the solutions for the
cantilever beam of length $L$ shown in Figure \ref{mensola} also
hold for a simply supported beam of length $2L$; this property will
be used in section \ref{sec:sec3}.

\subsection{Case (a): cantilever beam with eccentric axial load}\label{subseca}

Let us consider the cantilever beam in  figure \ref{mensola}
subjected to an axial force $N$ with eccentricity $e$ applied at the
free end. If we neglect the geometric nonlinearity, then the
curvature is constant along the axis, $\chi(x)=\bar\chi$, $x\in
[0,L]$, where $\bar\chi$ has the expression given by
\eqref{eq:eq5a}, with $M=N e$.


By integrating equation \eqref{eq:eq3} and imposing the boundary
conditions $y(0) = 0$, $y'(0) = 0$,  we obtain the transverse
displacement $y(x)$ of the beam
\begin{equation}
y(x)=-\frac{\bar\chi\,x^2}{2},\label{eq:eq6}
\end{equation}
and the displacement of the beam's upper end
\begin{equation}
f_a=-\frac{\bar\chi\,L^2}{2}.\label{eq:eq7a}
\end{equation}

Equation \eqref{eq:eq7a} reduces to the well known linear value
\begin{equation}\label{casea_linear}
\left|f_a\right|=\frac{\left|N\right|e}{2 E J}\,L^2
\end{equation}

\noindent for $e\le h/6$. When the eccentricity is outside the
kernel, the curvature is given by the second equation of
\eqref{eq:eq5a}, and using \eqref{eq:eq7a}, we obtain the
dimensionless expression

\begin{equation}
\frac{\left|f_a\right|}{L}=\frac{1}{9\left(
2e/h-1\right)^2}\frac{\left|N\right|}{\gamma},\label{eq:eq8}
\end{equation}
with

\begin{equation}
\gamma=3EJ/(L\,h).\label{eq:eq8a}
\end{equation}

\begin{figure}
\centering
\includegraphics[width=14cm]{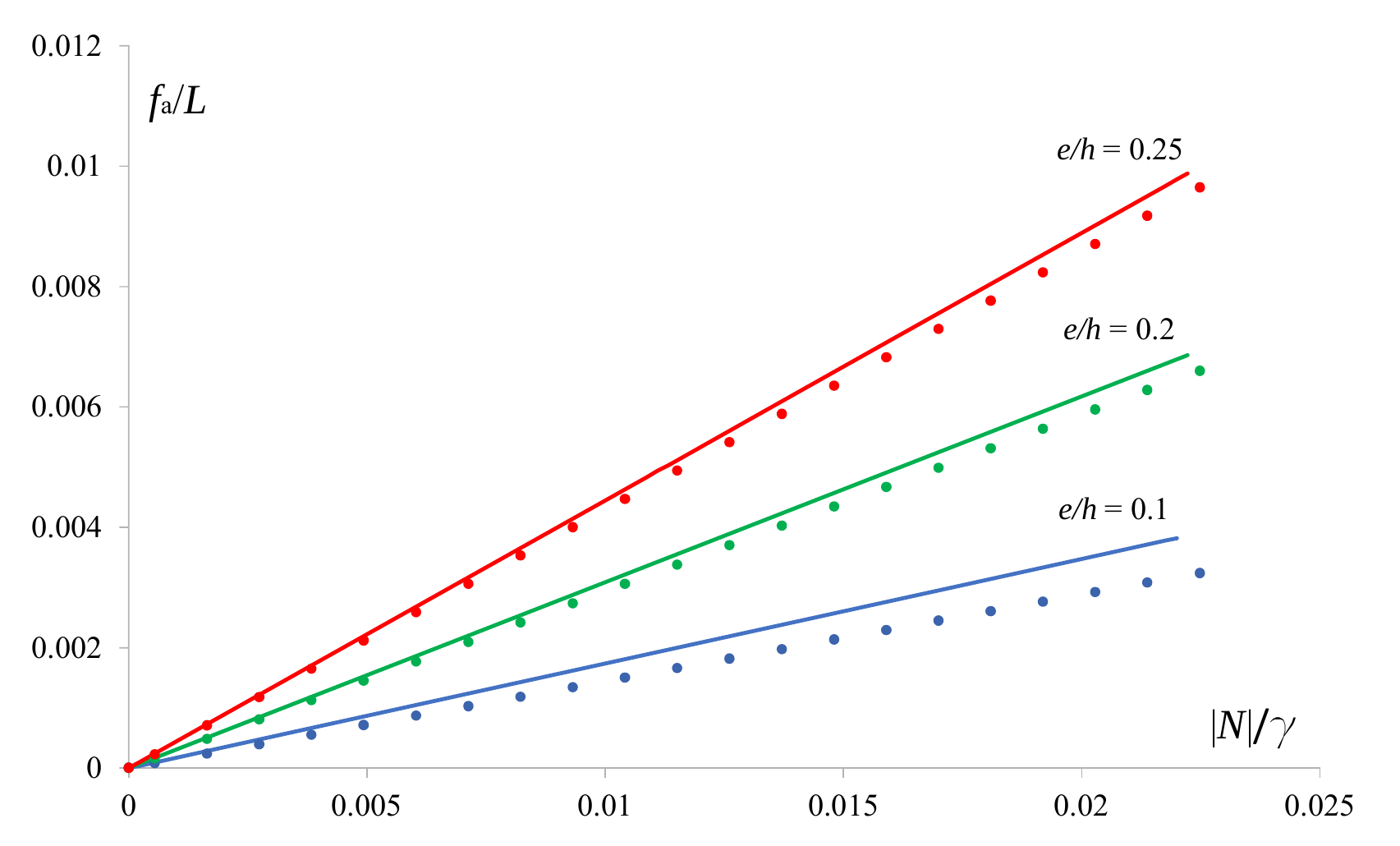}
\caption{Case (a): deflection of a masonry--like cantilever beam
$f_a/L$
vs normal force $|N|/\gamma$ for different values of the  eccentricity. 
}\label{Nfa_excel}
\end{figure}

\begin{figure}
\centering
\includegraphics[width=14cm]{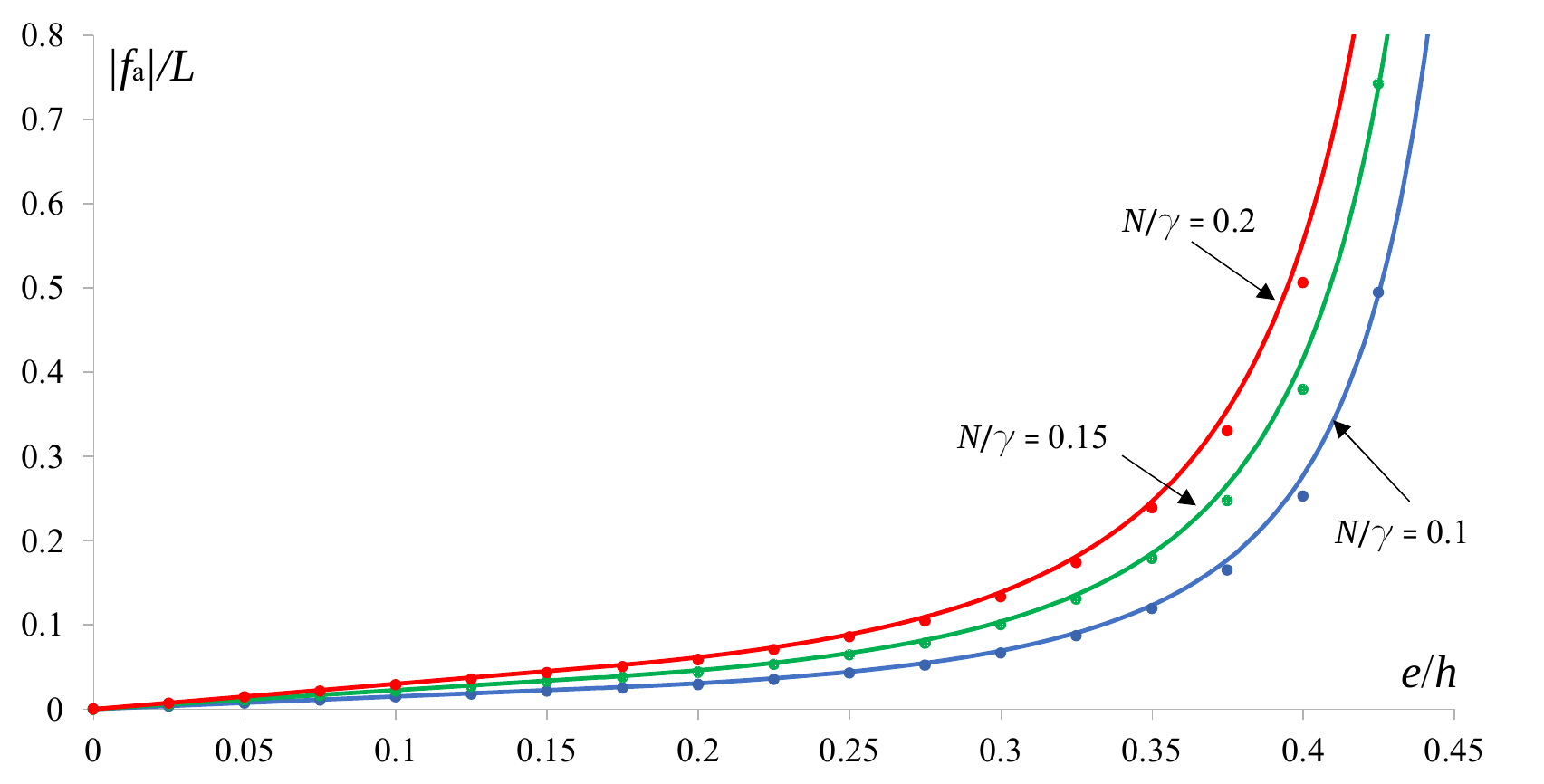}
\caption{Case (a): deflection $f_a$ of a masonry--like cantilever
beam vs the eccentricity $e$ acting on the beam, for different
values of the ratio $\left|N\right|/\gamma$.
}\label{struzzo_excel}
\end{figure}

Equations \eqref{casea_linear} and \eqref{eq:eq8} state that, when
the eccentricity remains fixed, there is a direct proportionality
between deflection and axial load, as shown in Figure
\ref{Nfa_excel}, plotting $|f_a|/L$ vs $|N|/\gamma$ for different
values of the eccentricity. In Figure \ref{struzzo_excel}, the
deflection of the beam is instead plotted against the eccentricity
$e$, for different values of the normal force. The figure highlights
the stiffness decay of the beam, when the eccentricity increases.

The analytical solutions have been compared with results obtained
via the Marc code \cite{marc}. In the numerical simulation, the
nonlinear concrete cracking model is chosen to simulate masonry,
with a tensile strength of $\unit{1\cdot 10^3}{\pascal}$ and a
Young's modulus $E=\unit{3\cdot 10^9}{\pascal}$, mass density
$\rho=\unit{1800}{\kilogram\per\metre^3}$, Poisson's ratio
$\nu=0.2$. The finite element model was built by using bi--linear
thick shell elements (element n. 75 of Marc library).

The results of the finite element analysis, represented by dotted
lines are reported in figures \ref{Nfa_excel} and
\ref{struzzo_excel}, along with the analytical counterparts.
Numerical and analytical simulations provide very similar results.
For small eccentricities, the numerical solution appears slightly
stiffer than the analytical one, because of the low but not null
tensile strength adopted in the constitutive equation available in
the Marc code.

\vspace{1\baselineskip}

If we take into account the second--order effects, while assuming
that \eqref{eq:eq3} still holds, we can write the beam's curvature
as follows

\begin{equation}
\label{eq:eq9} \left|\chi (x)\right|=
\begin{cases}
\quad \frac{\left|
N\right|}{EJ}\left(e+\left|y(L)-y(x)\right|\right) \quad &\text{for
$e+\left|y(L)-y(x)\right|
\le h/6$},\\
\vspace{1\baselineskip} \\
4 \,\alpha^3/\left(\frac{\left| N\right|}{EJ}
\left(e+\left|y(L)-y(x)\right|\right)-3\,\alpha\right)^2\quad
&\text{for\, $e+\left|y(L)-y(x)\right| > h/6$}.
\end{cases}
\end{equation}

As shown by equations \eqref{eq:eq9}, the curvature is no more
constant along the beam's axis. The differential equations
\eqref{eq:eq3}, \eqref{eq:eq9} can be numerically solved via the
following iterative scheme \cite{simmons}

\begin{equation}
\label{eq:eq11} \frac{d^2 y^{(n+1)}}{d x^2}=-\chi^{(n)},
\end{equation}

\begin{equation}
\label{eq:eq10} \left|\chi^{(n)}(x)\right|=
\begin{cases}
\quad \frac{\left|
N\right|}{EJ}\left(e+\left|y^{(n)}(L)-y^{(n)}(x)\right|\right) \quad
&\text{for $ e+\left|y^{(n)}(L)-y^{(n)}(x)\right|
\le h/6$},\\
\vspace{1\baselineskip} \\
4 \,\alpha^3/\left(\frac{\left| N\right|}{EJ}
\left(e+\left|y^{(n)}(L)-y^{(n)}(x)\right|\right)-3\,\alpha\right)^2\quad
&\text{for\, $e+\left|y^{(n)}(L)-y^{(n)}(x)\right| > h/6$},
\end{cases}
\end{equation}

and
\begin{equation}
\label{eq:eq12} y^{(0)}(x)=0,\,\,\,\,x\in [0,L].
\end{equation}

The algorithm converges when  the difference between deflection at
step $n+1$ and deflection at step $n$, $(n>0)$ falls under a fixed
residual value $\varepsilon$
\begin{equation}\label{residual}
\max\limits_{x\in[0,L]}\frac{\left|y^{(n+1)}(x)-y^{(n)}(x)\right|}{\left|y^{(n)}(x)\right|}<\varepsilon.
\end{equation}

The maximum eccentricity  along the beam's axis is
attained at $x=0$ and reaches, at the final step $n_{max}$, the
value
\begin{equation}\label{emax}
e_{max}=e+y^{(n_{max})}(L).
\end{equation}

\begin{figure}
\centering
\includegraphics[width=14cm]{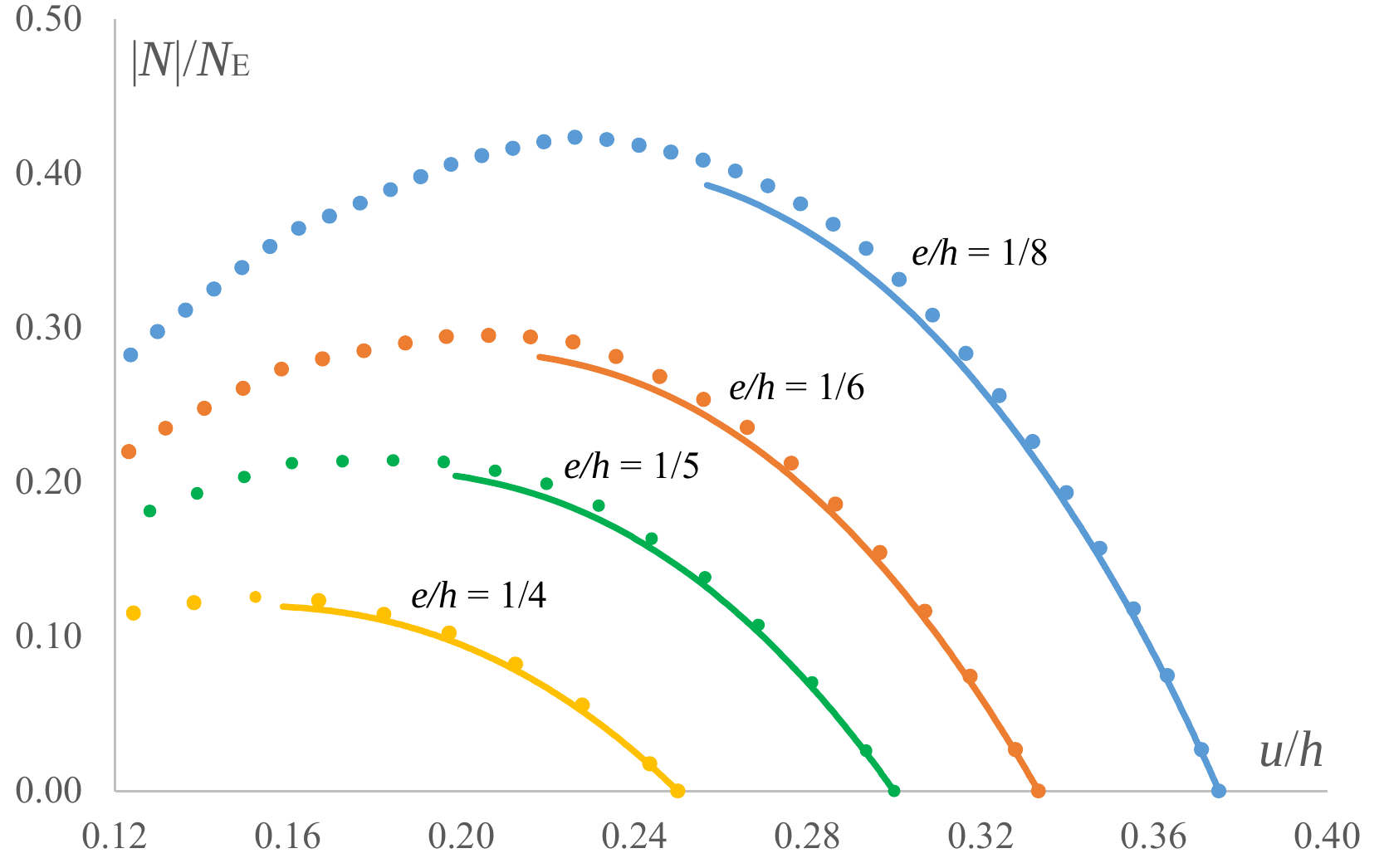}
\caption{Case (a) : ratio $\left|N\right|/N_E$ vs ratio $u/h$ for
different values of the initial eccentricity $e$. From the left:
$e/h=1/4$, $e/h=1/5$, $e/h=1/6$, $e/h=1/8$.}\label{Yokel}
\end{figure}

Let us define the quantities

\begin{equation}
\label{eq:eq13} N_E=\pi^2\frac{EJ}{(2L)^2},
\end{equation}

\begin{equation}
\label{eq:eq14} u=\frac{h}{2}-e_{max},
\end{equation}
which represent, respectively, the Euler's critical load of the
cantilever beam and the distance of the axial force from the
compressed edge at $x=0$. Ratio $\left| N\right|/N_E$ is plotted in
Figure \ref{Yokel} against ratio $u/h$, for different values of the
initial eccentricity $e$. If  $e$ exceeds the limit elastic value
$h/6$, the whole process takes place in the nonlinear field, while,
for $e\le h/6$, the beam, which initially behaves linearly, cracks
during the process due to the geometric nonlinearity. The iteration
scheme has been implemented in the \emph{Mathematica} environment
\cite{Mathematica} and calculations are performed by setting the
residual $\varepsilon$ (in percentage) at $0.001$ and allowing the
maximum number of thirty iterations. The solutions, shown in the
Figure \ref{Yokel} via solid lines, coincide with the explicit
curves reported in \cite{ADF2003}, \cite{yokel1971}. The results of
the finite element analysis conducted via the Marc code (dots) are
also reported in the figure and show an excellent agreement with
those of algorithm \eqref{eq:eq11}--\eqref{eq:eq12}. The finite
element analysis exploits the arc--length method \cite{marc} to
evaluate the beam's behaviour after buckling, while the algorithm is
not able to follow the post--critical behaviour of the system.

\begin{figure}
\centering
\includegraphics[width=14cm]{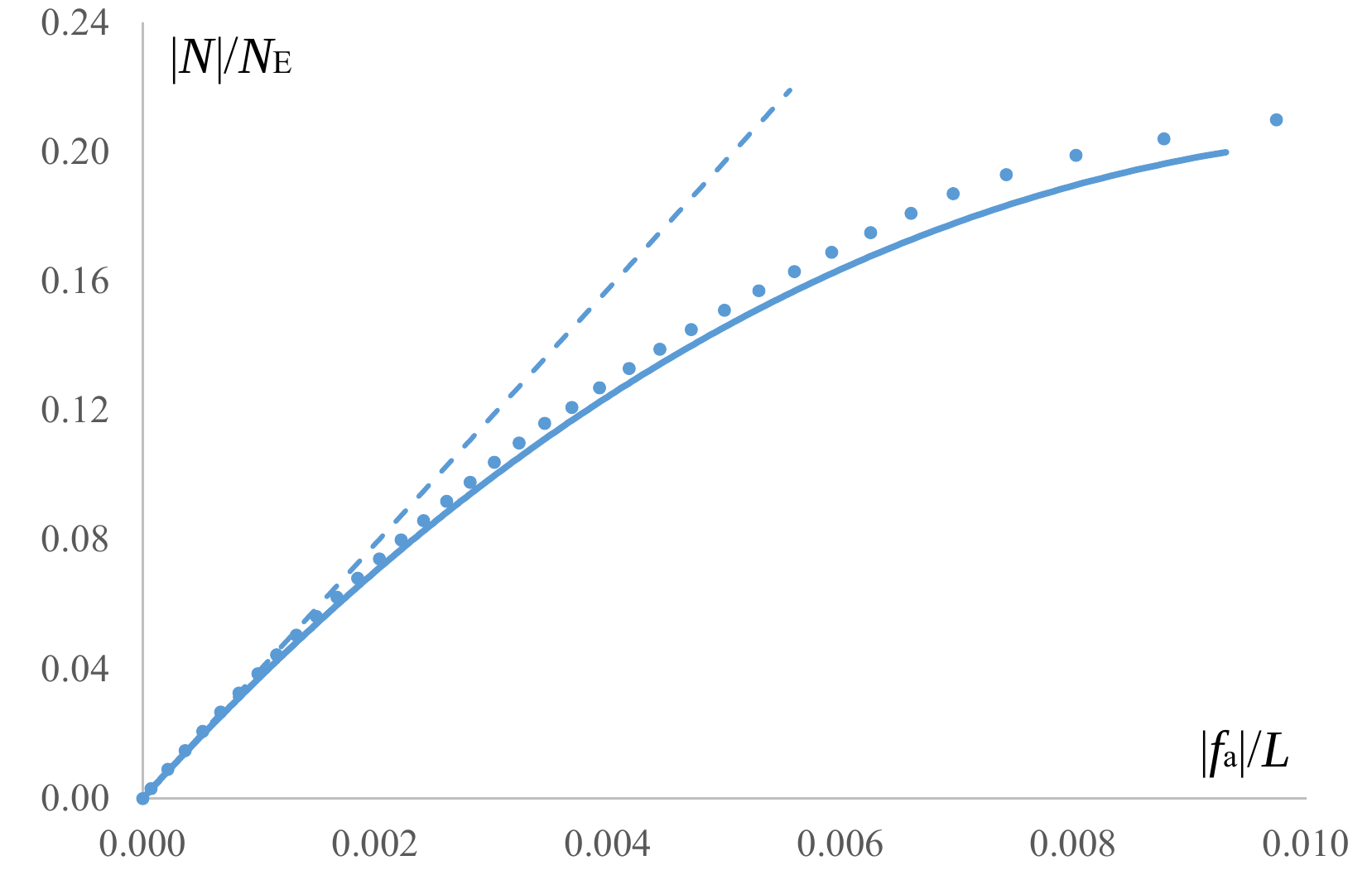}
\caption{Case (a): load--displacement curves for $e/h=1/5$.
Masonry--like material (dashed curve); masonry--like material with
geometric nonlinearity (continuous curve); finite--element
simulation (dots).}\label{pdelta}
\end{figure}

Figure \ref{pdelta} shows the dimensionless load--displacement
curves for the masonry--like case (dashed line), and under the
combined effect of the constitutive and geometric nonlinearity
(continuous line). Dots represent the results of the numerical
simulation performed via Marc code. The curves depend on the value
of the ratio between the eccentricity $e$ at the top of the column
and the height $h$ of the section, which is here set to $1/5$. The
abscissa $\left| f_a\right|/L$ represents the deflection of the
beam's top to the beam's length. The effects of cracking on the
stiffness of the beam are evident. However, considering only the
constitutive nonlinearity, it does not allow modelling the collapse,
which is captured instead  when the effects of deformation are taken
into account in the equilibrium equations.

\subsection{Case (b): cantilever beam with axial and horizontal loads}\label{subsecb}

Let us consider the cantilever beam subjected to an axial force $N$
and a horizontal force $H$,  applied to the upper end. The
eccentricity of the normal force is no more constant along the
beam's length; in fact, in the case of small deflection the
first--order bending moment $M$ has the expression

\begin{equation}
\label{eq:eq15} M(x)=H(L-x).
\end{equation}

The horizontal force $H$, supposed nonnegative, ranges in the
interval $[0, H_{max}]$, where

\begin{equation}\label{hmax}
H_{max}=\frac{\left|N \right|h}{2L}
\end{equation}

\noindent is the load corresponding to which the normal force $N$ is
applied at the edge of the base section.

Thus, let us define the abscissa $x_0$ as the position along the
beam's length in which the curvature takes the limit elastic value
$\alpha$ given in  \eqref{eq:eq2}, that is

\begin{align}
\label{eq:eq16} \left|\chi(x_0)\right|&=\alpha,\\
\label{eq:eq16b}\left|M(x_0)\right|&=\left|N\right|h/6.
\end{align}

Equations \eqref{eq:eq15} and \eqref{eq:eq16b}  give

\begin{equation}
\label{eq:eq16bis}x_0=L-\frac{\left|N\right| h}{6 H}=
L-\frac{\alpha}{k},
\end{equation}

\noindent with

\begin{equation}
\label{eq:eqk}k=\frac{H}{EJ}.
\end{equation}

If the beam is all in the linear elastic field, that is, if

\begin{equation}
H\le\frac{\left|N \right| h}{6L},
\end{equation}

\noindent then $x_0$ is zero, and the solution to equation
\eqref{eq:eq3} coincides with the linear elastic deflection.

For $x_0>0$, from \eqref{eq:eq5a}, we get the expression of the
curvature $\chi$ along the axis

\begin{equation}
\label{eq:eq17}  \chi(x)=
\begin{cases}
\quad \frac{H(L-x)}{EJ} \quad &\text{for $ x
\ge x_0$},\\
\vspace{1\baselineskip} \\
4 \,\alpha^3/\left(\frac{ H(L-x)}{EJ}-3\,\alpha\right)^2\quad
&\text{for\, $x < x_0$}.
\end{cases}
\end{equation}

Therefore, the solution of equation \eqref{eq:eq3} splits  into two
parts, $y=y_1(x)$ for $x\le x_0$, $y=y_2(x)$ for $x>x_0$, where
$y_1$ and $y_2$ have the following explicit expressions,

\begin{equation}
\label{eq:eq18}y_1(x)=c_1+c_2
x+\frac{4\alpha^3}{k^2}\,\log\big(k(x-L)+3\alpha\big) \quad
\text{for $ x \le x_0$},
\end{equation}

\begin{equation}
\label{eq:eq18}y_2=c_3+c_4 x-\frac{k L}{2} x^2+\frac{k}{6} x^3,
\quad \text{for\, $x > x_0$},
\end{equation}

\noindent with $k$ given by \eqref{eq:eqk}.
The conditions

\begin{equation}
\begin{cases}
\label{eq:eq19}  y_1(x_0)=y_2(x_0),\\
\vspace{1\baselineskip} \\
y'_1(x_0)=y'_2(x_0),\\
\vspace{1\baselineskip} \\
y_1(0)=y'_1(0)=0.\\
\end{cases}
\end{equation}

\noindent allow determining the constants

\begin{align}
& c_1=-\frac{4\alpha^3}{k^2}\log(3\alpha-kL),\label{eq:eq20}\\
& c_2=-\frac{4\alpha^3}{k}\frac{1}{(3\alpha-kL)},\label{eq:eq21}\\
& c_3=-\frac{(kL-\alpha)^3}{2k(3\alpha-kL)},\label{eq:eq22}\\
&
c_4=-\frac{1}{6k^2}\left[k^3L^3+9kL\alpha^2-10\alpha^3-24\alpha^3\big(\log{2\alpha}-\log{(3\alpha-k
L)}\big)\right].\label{eq:eq23}
\end{align}

\vspace{1\baselineskip}

 The deflection of the top of the cantilever beam takes thus the expression

\begin{equation}
\label{eq:eq23} f_a= -\frac{\alpha^3}{3k^2(3\alpha-k L)} \left\{ 17k
L-15\alpha-12(3\alpha-k L)\left[\log{2\alpha}-\log{(3\alpha-k
L)}\right]\right\},
\end{equation}

\noindent and the problem is governed by the curvature

\begin{equation}
\label{eq:eq24} \zeta= 3\alpha-k L.
\end{equation}

It is an easy matter to prove that
\begin{equation}\label{eq:eq24b}
\zeta=3\alpha(1-\left|M(0)\right|/M_{max}),
\end{equation}

\noindent with $M(0)$ the bending moment at the beam's base, and
\begin{equation}\label{eq:eq24b}
M_{max}=H_{max} \,L
\end{equation}
\noindent the maximum bending moment sustainable by the
beam's section.

The curvature $\zeta$ tends to zero when $\left|M(0)\right|$ tends
to $M_{max}$. Correspondingly, the deflection \eqref{eq:eq23} of the
beam tends to infinity. When $\left|M(0)\right|$ tends to the limit
linear value $\left|N\right|h/6$, then $ k=\alpha/L$ and thus
$\zeta$ tends to $2\alpha$. In this case  the deflection of the beam
coincides with the well known  value $\left|f_a\right|= k L^3/3$.

By introducing the quantities

\begin{equation}
\label{eq:eq25} \bar k=kL^2,\,\,\,\,\,\,\,\bar\alpha=\alpha
L,\,\,\,\,\,\,\,\, \bar\zeta=\zeta L,
\end{equation}

\noindent equation \eqref{eq:eq23} takes the dimensionless form

\begin{equation}
\label{eq:eq23b} \frac{f_a}{L}= -\frac{\bar\alpha^3}{3\bar k^2
\bar\zeta} \left(17\,\bar
k-15\,\bar\alpha-12\,\bar\zeta\log{\frac{2\bar\alpha}{\bar\zeta}}\right).
\end{equation}

\begin{figure}
\centering
\includegraphics[width=14cm]{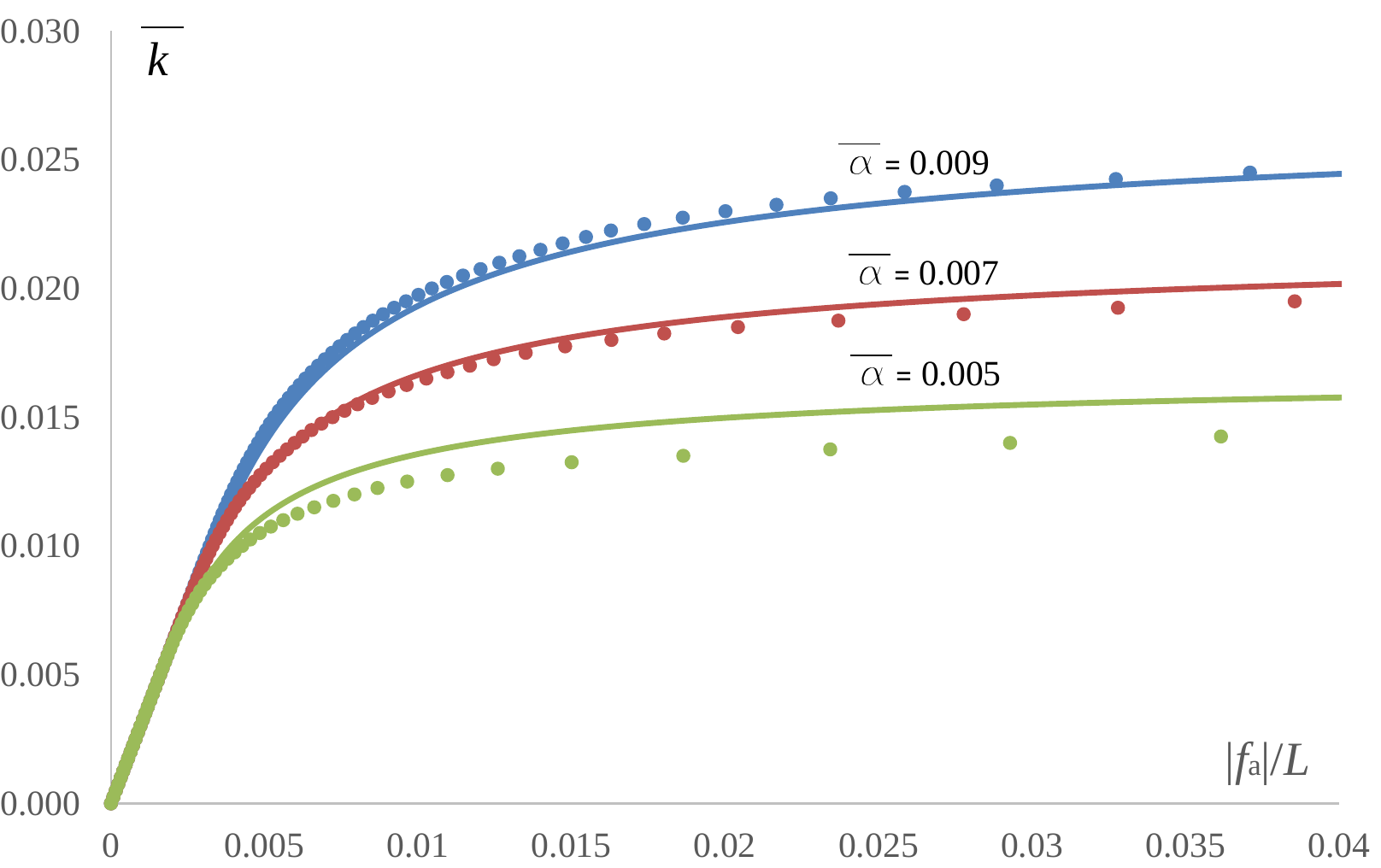}
\caption{Case (b): push--over curves for different values of
$\bar\alpha$. Dots represent the finite--element
solution.}\label{hdelta}
\end{figure}

In figure \ref{hdelta} some  load--displacement curves are shown
(push--over curves) for different values of the normal force acting
on the beam's section (measured by the dimensionless limit elastic
curvature $\bar\alpha$).

\vspace{1\baselineskip}

As done for case (a), to take into account the geometric
nonlinearity we can adopt the following iterative scheme

\begin{equation}
\label{eq:eq24alg} \frac{d^2 y^{(n+1)}}{dx^2}=-\chi^{(n)},
\end{equation}

\begin{equation}
\label{eq:eq25} \left|\chi^{(n)}(x)\right|=
\begin{cases}
\quad \frac{\left| H\right|}{EJ}(L-x)+\frac{\left|
N\right|}{EJ}\left|y^{(n)}(L)-y^{(n)}(x)\right| \quad &\text{for $ x
\ge x_0^{(n)}$},\\
\vspace{1\baselineskip} \\
4 \,\alpha^3/\left(\frac{\left| H\right|}{EJ}(L-x)+\frac{\left|
N\right|}{EJ}
\left|y^{(n)}(L)-y^{(n)}(x)\right|-3\,\alpha\right)^2\quad
&\text{for\, $x < x_0^{(n)}$},
\end{cases}
\end{equation}

\begin{equation}
\label{eq:eq26} y^{(0)}(x)=0,
\end{equation}

\noindent $x_0^{(0)}$ is given by \eqref{eq:eq16bis}, and
$x^{(n)}_0$ is root of the equation
\begin{equation}\label{x0n}
\left|H\right|(L-x)+|N|\left|y^{(n)}(L)-y^{(n)}(x)\right| =
\frac{|N| h}{6}.
\end{equation}

As for the previous case, iterations \eqref{eq:eq24alg}--
\eqref{x0n} are repeated  until the difference between two
consecutive deflections becomes smaller than a prescribed residual
value $\varepsilon$.

\begin{figure}[h]
\centering
\includegraphics[width=14cm]{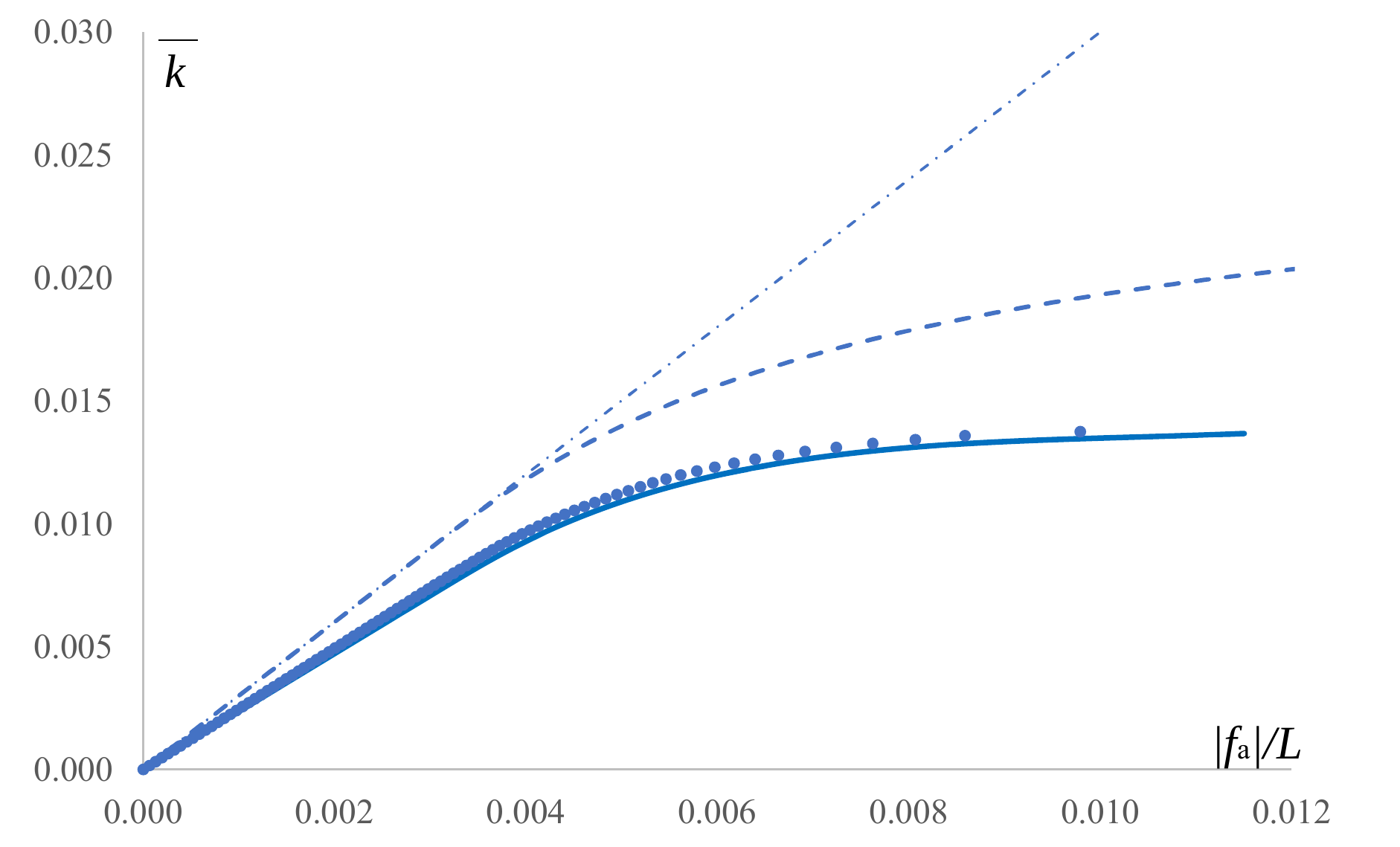}
\caption{Case (b): push--over curves for $\bar\alpha=9\cdot10^{-3}$.
linear elastic solution (dash--dotted line); masonry--like solution
(dashed line); masonry--like solution with geometric nonlinearity
(continuous line); finite--element solution
(dots).}\label{hdelta_80000}
\end{figure}

\begin{figure}[h]
\centering
\includegraphics[width=12cm]{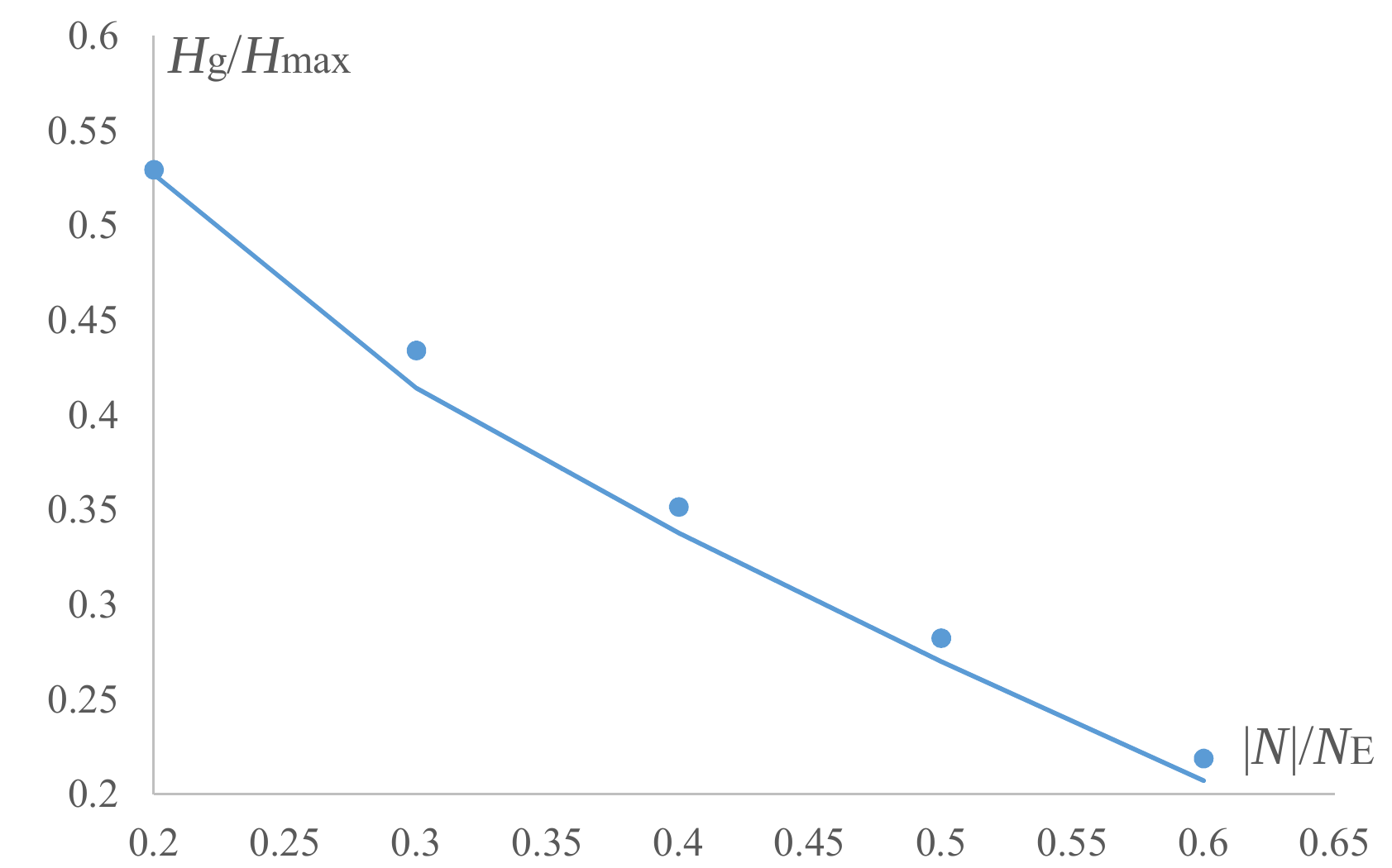}
\caption{Case (b): ratio between the collapse load $H_g$ calculated
by taking into account geometric nonlinearity and  $H_{max}$, vs the
ratio $\left|N\right|/N_E$, analytical results (continuous line) and
finite--element results (dots).}\label{Hred}
\end{figure}

Figure \ref{hdelta_80000} shows the push--over curves for
$\bar\alpha=9\cdot10^{-3}$ in the linear elastic case (dash-dotted
line), in the masonry--like case (dashed line), and in the
masonry--like case  with geometric nonlinearity (continuous line),
via algorithm \eqref{eq:eq24alg}--\eqref{eq:eq26}. Calculations has
been performed setting the residual $\varepsilon$ (percentage) to
0.001 and the maximum number of iterations to 30. Dots represent the
results of the finite--element analysis. Second--order effects cause
a significant change in the system's response to the horizontal
load; the collapse load $H_g$ is about one half of the maximum
horizontal load $H_{max}$ admissible in the masonry--like solution.
Figure \ref{Hred} shows the ratio between $H_g$ and $H_{max}$ vs the
axial force $N$ (in the abscissa the ratio $\left|N\right|/N_E$).
The figure highlights the reduction of the beam's strength when the
geometric nonlinearity is taken into account. Results of the
finite--element analysis are also reported in the figure (dots). As
a results of the different model adopted, the collapse load
evaluated by the Marc code is always slightly larger than that
evaluated via the masonry--like constitutive equation.

\section{Influence of  geometric nonlinearity on the natural frequencies of masonry beams}\label{sec:sec3}

The results shown in the previous section can be used to evaluate
the effects of deflections on the natural frequencies of masonry
beams, by using the method proposed in \cite{Girardi}, where the
fundamental frequency of a simply supported masonry beam subjected
to some load conditions is calculated explicitly.

\begin{figure}[h]
\centering
\includegraphics[width=12cm]{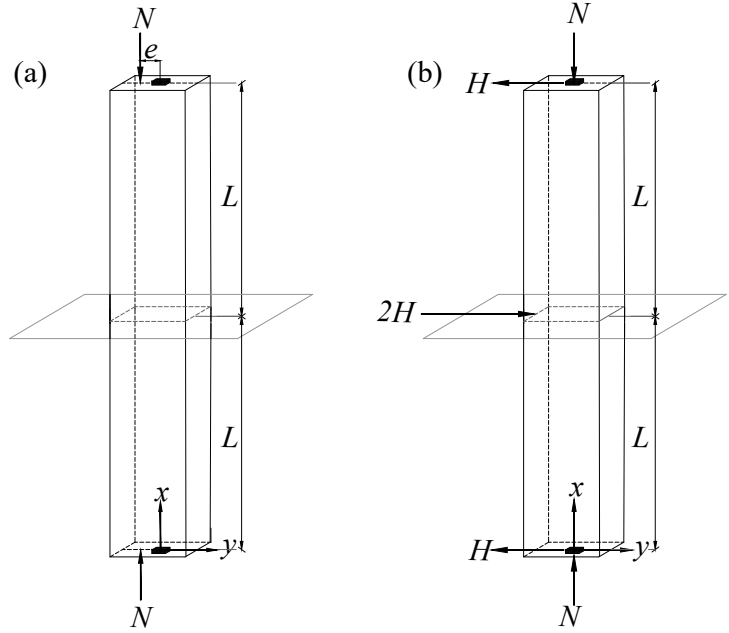}
\caption{Simply supported scheme corresponding to cases (a) and
(b).}\label{app_freq}
\end{figure}

The geometry of the beam is described in figure \ref{app_freq}, for
both cases (a) and (b).


By taking into account the axial force $N$, the motion of the beam
is governed by the equation

\begin{equation}
y_{t t}-(f(\chi))_{x x}=\left[p(x,t)+ N \, y_{xx}
\right]\frac{1}{\rho b h} \label{eq:eq3add},
\end{equation}
where  $ x \in [0,2L]$ is the abscissa along the beam's axis (figure
\ref{app_freq}), $p(x,t)$ is the transverse load per unit length,
and

\begin{equation}
f(\chi)=\frac{M(\chi)}{\rho\,b\,h}.\label{eq:eq2add}
\end{equation}

Let us consider, at $t=0$, a load $\bar p(x)$ inducing in the beam
an initial deflection $\bar y(x)$ and curvature change
$\bar\chi(x)$. The beam reaches the equilibrium under load $\bar p$,
and thus

\begin{equation}
-\left(f(\bar\chi)\right)_{ x x}=\left[\bar p(x)-
N\,\bar\chi\right]\frac{1}{\rho b h}\label{eq:eq3badd}.
\end{equation}

We are interested in studying the small oscillations $\delta y$ of
the beam around $\bar y $

\begin{equation}
\delta y_{t t}-\left(\frac{d f}{d\chi}
\Bigg\vert_{\bar\chi}\delta\chi\right)_{x  x}=-\frac{N}{\rho b
h}\,\delta\chi \label{eq:eq6add},
\end{equation}
where we used the approximation

\begin{equation}
f(\bar\chi+\delta\chi)\simeq f(\bar\chi)+\frac{d
f}{d\chi} \Bigg\vert_{\bar\chi}\delta\chi, \label{eq:eq5}
\end{equation}
and condition \eqref{eq:eq3badd}. We assume the small oscillations
$\delta y$ have the approximate expression

\begin{equation}
\delta y\simeq \delta a \sin\left(\frac{\pi}{2L}  x\right) u(t),
\label{eq:eq7}
\end{equation}
with $\delta a>0$. By following the procedure described in
\cite{Girardi}, from \eqref{eq:eq6add} we get an approximation of
the fundamental frequency

\begin{equation}
\omega^2 \simeq
\frac{2\pi^4}{(2L)^5}\int_{0}^{2L}\sin^2\left(\frac{\pi}{2L}
 x\right)\frac{d f}{d\chi} \Bigg\vert_{\bar\chi}
dx+\frac{c^2 \pi^4}{(2L)^4}\frac{N}{N_E}, \label{eq:eq9add}
\end{equation}
where

\begin{equation}
\label{eq:eq12add} \frac{d f}{d \chi}\Bigg\vert_{\bar\chi}=
\begin{cases}
\quad c^2  \quad &\text{for $\left| \bar \chi \right|
\le \alpha$},\\
\quad  c^2 \sqrt{\frac{\alpha^3}{{\left| \bar\chi \right|}^3}}\quad
&\text{for} \left| \bar\chi \right|
> \alpha,
\end{cases}
\end{equation}

\noindent  $c$ is the elastic constant of the beam
\begin{equation}
c^2= \frac{E J}{\rho b h},
\end{equation}

\noindent and $N_E$ is the Euler load expressed by \eqref{eq:eq13}.

If the material constituting the beam is linear elastic, by using
the first equation of \eqref{eq:eq12add}, equation \eqref{eq:eq9add}
becomes

\begin{equation}
\omega^2 = \omega_{el}^2\left(1-\frac{\left|N\right|}{N_E}\right), \label{eq:eq13add}
\end{equation}

\noindent with
\begin{equation}
\omega_{el}^2= \frac{c^2 \pi^4}{(2 L)^4}.
\end{equation}

In the case of masonry--like material, first term in
\eqref{eq:eq9add} takes into account both constitutive and geometric
nonlinearity in the equilibrium equation \eqref{eq:eq3badd}; the
fundamental frequency $\omega^2$ can be calculated by using function
$\bar\chi$ determined via the algorithm shown in the previous
sections. 

%
%

\subsection{Case (a): simply supported beam with eccentric axial load}\label{subsecadin}

Let us consider the beam represented in figure \ref{app_freq}. The
beam is subjected to a constant axial force $N$  with eccentricity
$e$. The deformation of the beam  can be obtained form that of the
cantilever beam of length $L$ (Case (a), subsection \ref{subseca}).
Thus, equation \eqref{eq:eq9add} becomes

\begin{equation}
\label{eq:eq28} \omega^2 \simeq  \frac{4\pi^4 c^2}{(2
L)^5}\int_0^{L} \sin^2\left(\frac{\pi x}{ 2
L}\right)\sqrt{\frac{\alpha^3}{\left|\bar\chi\right|^3}}\,dx
+\frac{c^2 \pi^4}{(2L)^4}\frac{N}{N_E}\quad \text{for} \left|
\bar\chi \right|
> \alpha,
\end{equation}

\noindent with $\left|\bar\chi\right|$ given by
%
%
algorithm \eqref{eq:eq11}--\eqref{eq:eq12}.

\begin{figure}[h]
\centering
\includegraphics[width=14cm]{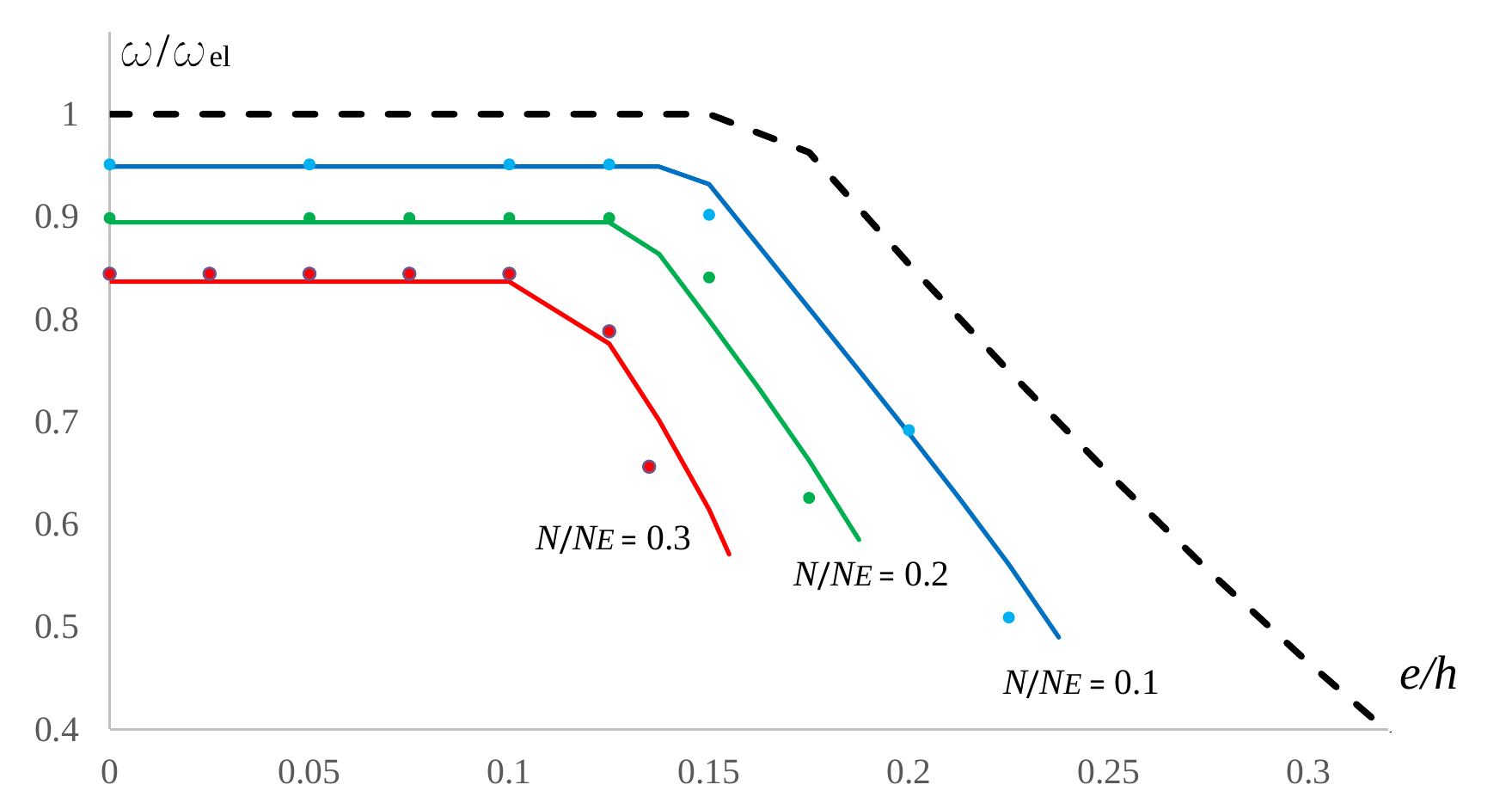}
\caption{Case (a). Ratio $\omega/\omega_{el}$ vs  $e/h$ for a
masonry--like beam by taking into account the geometric
nonlinearity.  $\left|N\right|/N_E=0.3$ (red curve);
$\left|N\right|/N_E=0.2$ (green curve); $\left|N\right|/N_E=0.1$
(blue curve);  finite--element simulation (dots); masonry--like
solution without geometric nonlinearity (dashed black line).}
\label{freq_ng}
\end{figure}

Figure \ref{freq_ng} shows ratio $\omega/\omega_{el}$ vs ratio
$e/h$, for different values of the normal force $\left|N\right|/N_E$
acting on the beam, and $N_E$ the Euler critical load given by
\eqref{eq:eq13}. The black dashed curve represents the masonry--like
solution without taking into account geometric nonlinearity. This
last solution, as shown in \cite{Girardi}, does not depend on the
normal force. On the contrary, taking into account the effects of
deflections induces a strong dependence of the solution on the
normal force. In particular, as shown by Figure \ref{freq_ng}, when
the ratio $\left|N\right|/N_E$ increases the beam's fundamental
frequency quickly decreases. For $\left|N\right|/N_E = 0.3$ (red
line in the figure), the beam approaches the collapse when the load
is applied with an eccentricity of only $1/6$ of the section's
height $h$; the corresponding  solution without taking into account
the geometric nonlinearity (black dashed curve) remains entirely in
the linear elastic field, and the fundamental frequency coincides
with $\omega_{el}$.

In Figure \ref{freq_ng} the dots represent the results of the
finite--element simulation, obtained via the prestressed modal
analysis procedure \cite{marc}. The figure shows a very good
agreement between the results obtained via equation \eqref{eq:eq28}
and those evaluated by the finite--element code.

\subsection{Case (b): simply supported beam with axial and horizontal loads}\label{subsecbdin}

Case \textit{b} considers a simply supported beam of length $2L$
subjected to a concentrated load $2H$ in the mid--section, as shown
in figure \ref{app_freq}. For this structure the equivalence applies
to Case (b) of subsection \ref{subsecb}.

The fundamental frequency of the beam can be deduced from
\eqref{eq:eq9add}

\begin{equation}
\label{eq:eq30} \omega^2=  \frac{4\pi^4 c^2}{(2
L)^5}\left(\int_0^{x_0} \sin^2\left(\frac{\pi x}{2
L}\right)\sqrt{\frac{\alpha^3}{|\bar\chi|^3}}\,dx +
\int_{x_0}^{L}\sin^2\left(\frac{\pi x}{2
L}\right)\,dx\right)+\frac{c^2 \pi^4}{(2L)^4}\frac{N}{N_E}
\end{equation}

\noindent with $x_0$ the abscissa of the beam's section in which
$\left|\bar \chi\right|=\alpha$, and
 $\left|\bar\chi\right|$ determined  via
algorithm \eqref{eq:eq24alg}--\eqref{x0n}.

\begin{figure}[h]
\centering
\includegraphics[width=12cm]{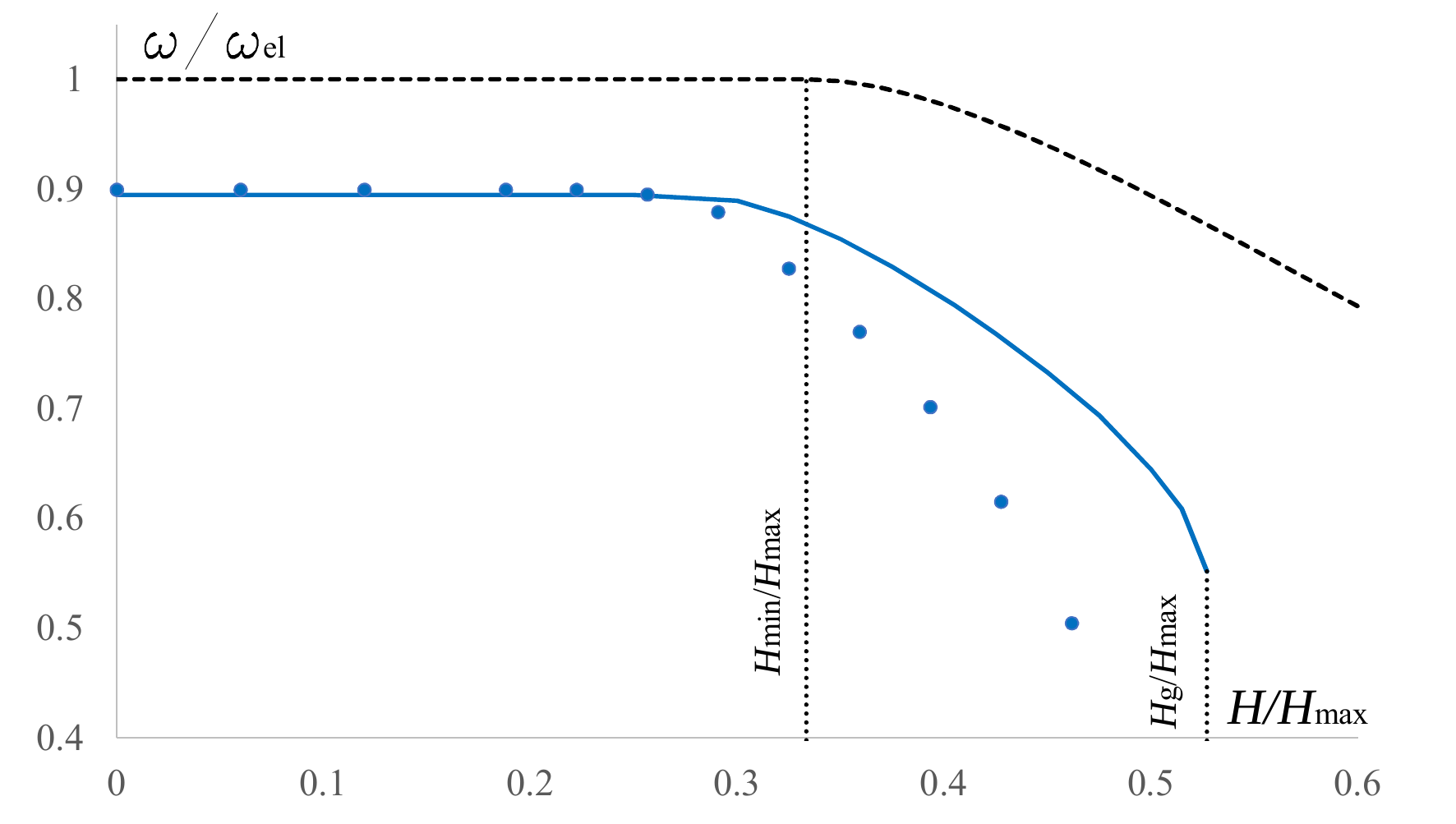}
\caption{Case (b): ratio $\omega/\omega_{el}$ vs  $H/H_{max}$ for a
masonry--like beam and $\left|N\right|/N_E=0.2$. Masonry--like
without geometric nonlinearity (dashed line);  masonry--like by
taking into account geometric nonlinearity (continuous line);
finite--element simulation (dots).} \label{freq_conc_ng02}
\end{figure}

\begin{figure}[h]
\centering
\includegraphics[width=12cm]{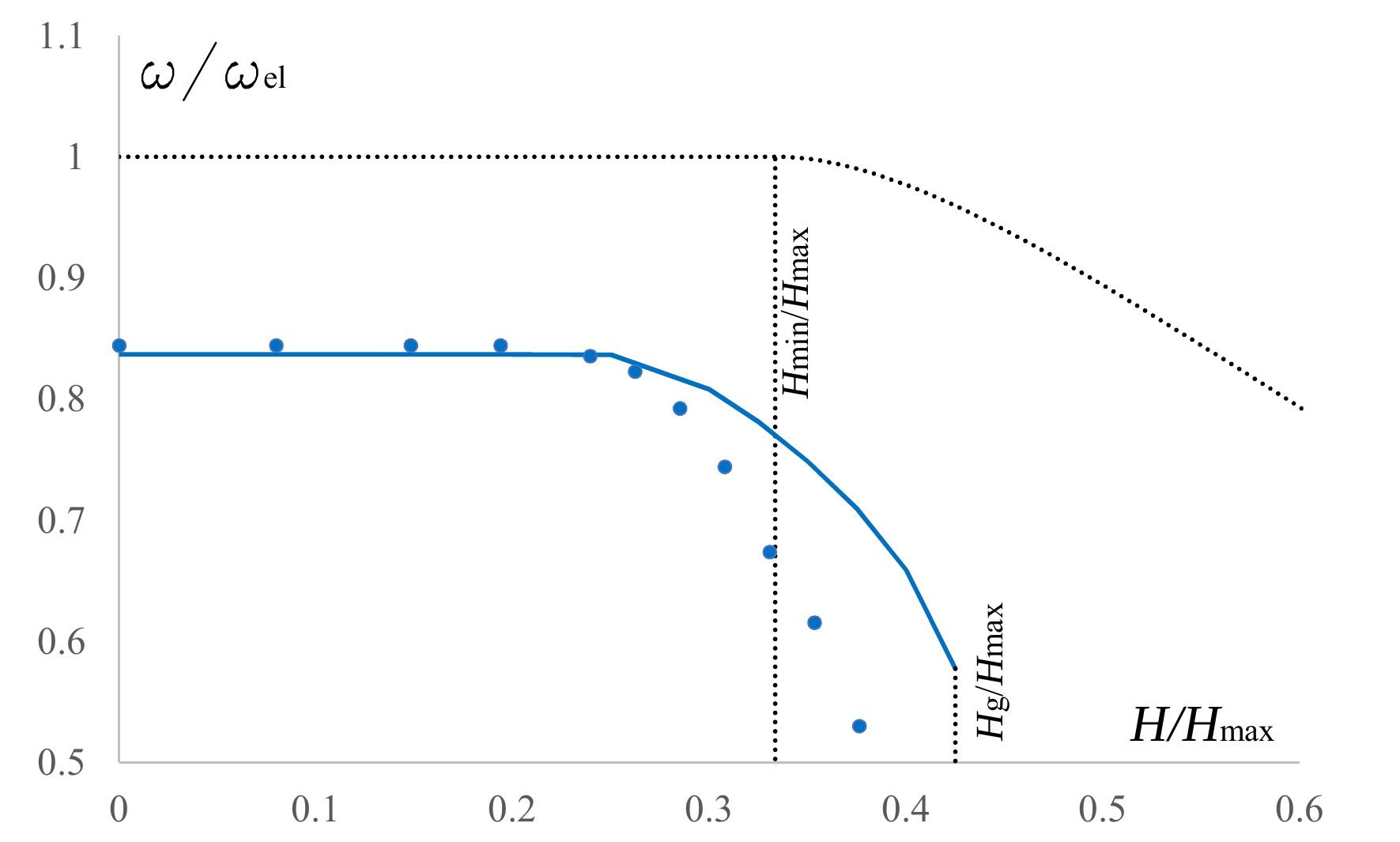}
\caption{Case (b): ratio $\omega/\omega_{el}$ vs  $H/H_{max}$ for a
masonry--like beam and $\left|N\right|/N_E=0.3$.  Masonry--like
without geometric nonlinearity (dashed line);  masonry--like by
taking into account geometric nonlinearity (continuous line);
finite--element simulation (dots).} \label{freq_conc_ng03}
\end{figure}

\begin{figure}[h]
\centering
\includegraphics[width=12cm]{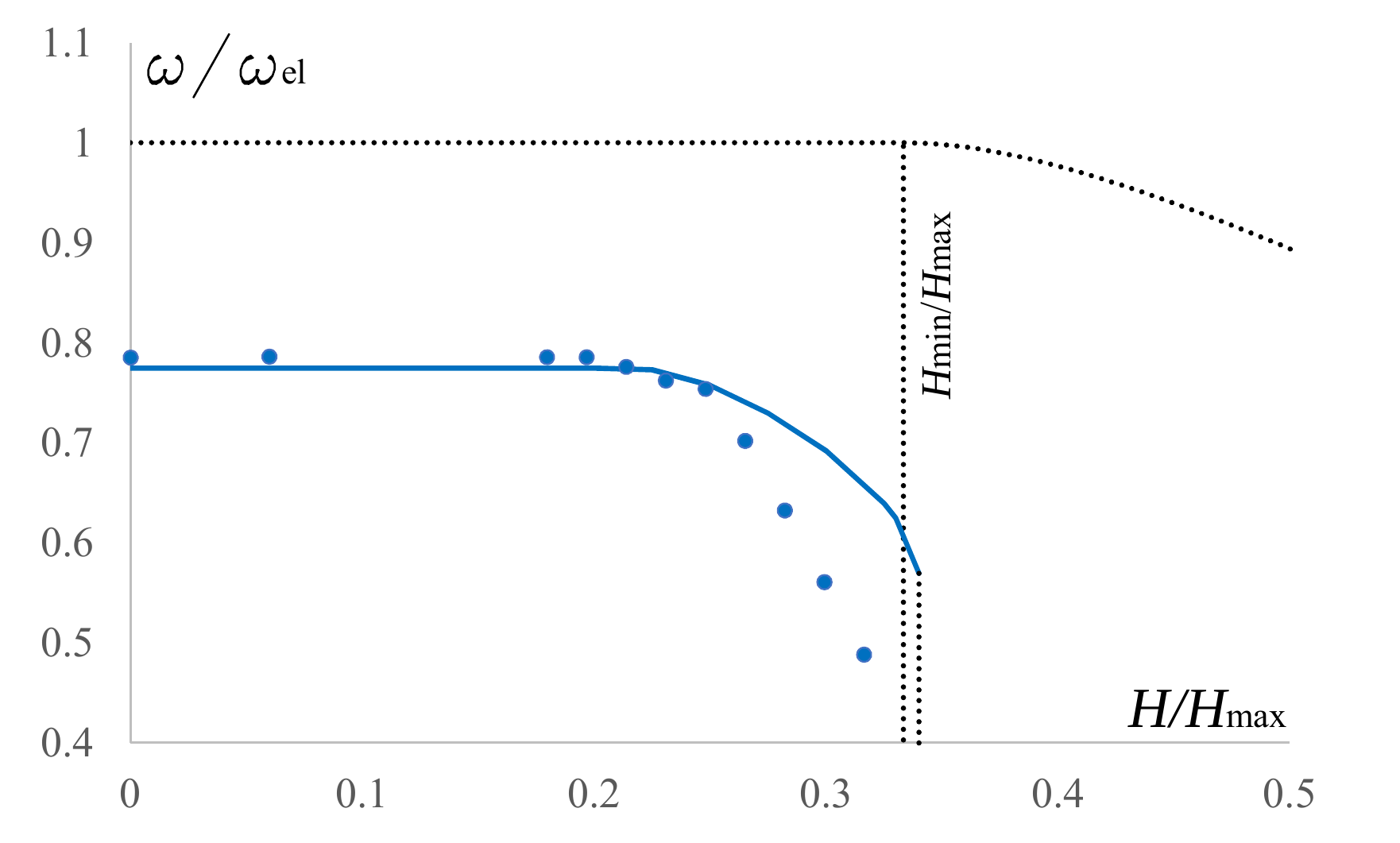}
\caption{Case (b): ratio $\omega/\omega_{el}$ vs  $H/H_{max}$ for a
masonry--like beam and $\left|N\right|/N_E=0.4$.  Masonry--like
without geometric nonlinearity (dashed line);  masonry--like by
taking into account geometric nonlinearity (continuous line);
finite--element simulation (dots).} \label{freq_conc_ng04}
\end{figure}

Figures \ref{freq_conc_ng02} to \ref{freq_conc_ng04} show the
fundamental frequency $\omega/\omega{el}$ of the beam vs the
horizontal force acting (ratio $H/H_{max}$), for different values of
$N/N_E$ (ranging from 0.2 in Figure \ref{freq_conc_ng02} to 0.4 in
Figure \ref{freq_conc_ng04}). The dashed line is for the
masonry--like material without geometric nonlinearity: the
fundamental frequency of the beam is equal to $\omega_{el}$  for
$H\le H_{min}$ and decreases for greater values of $H$. The
continuous  lines show the frequency of the beam when the geometric
nonlinearity is also taken into account. In this case, the frequency
 begins to decrease when the horizontal force is lower
than $H_{min}$  and falls down when the structures reaches the
collapse load $H_g$ (see Figure \ref{Hred}). For $N=0.4\, N_E$ the
collapse load $H_g$ is essentially equal to $H_{min}$ and, thus, the
geometric nonlinearity cuts the frequency while the corresponding
masonry--like curve is still in the linear field. The figures
present also the results of the finite--element simulation (dots)
conducted via the prestressed modal analysis implemented in
\cite{marc}.


%
%


\section*{Conclusions}\label{sec:sec4}

The present paper investigates the influence of geometric
nonlinearity on the static and dynamic behaviour of Euler- Bernoulli
beams made of a masonry-like material. In the first part, the static
behaviour of a cantilever beam subjected to an eccentric normal load
(case (a)) and to axial and horizontal loads (case (b)) is
addressed. The knowledge of the normal force and bending moment
along the beam's axis makes it possible to calculate the deflection
while considering both material and geometric nonlinearities. The
nonlinear differential equation that links deflection and curvature
when second-order effects are taken into account is integrated via
an iterative scheme. Several response curves for case (a) and
push-over curves for case (b) are reported to highlight how
geometric nonlinearity reduces the static performance of the beam.
The second part of the paper aims to assess the influence of
geometric nonlinearity on the fundamental frequency of a simply
supported masonry beam. In particular, the frequency is explicitly
calculated exploiting the knowledge of the beam's deflection for the
cases (a) and (b) dealt with in the first part. It is worth noting
that the reduction of the fundamental frequency due to the presence
of cracks is remarkably exacerbated by the geometric nonlinearity.
The results concerning the static and dynamic behaviour of the beam
are corroborated by finite-element analysis.

\end{document}